\documentclass{PoS}
\pdfoutput=1

\usepackage[utf8]{inputenc}
\usepackage{tikz}
\usetikzlibrary{shapes,arrows}

\tikzstyle{block}=[rectangle, draw, fill=blue!20, text width=9.5em, 
                   text centered, rounded corners, minimum height=2em, 
                   minimum width=10em]
\tikzstyle{line} =[draw, -latex']

\title{The PDFLattice2017 workshop: a summary report}

\ShortTitle{The PDFLattice2017 workshop: a summary report}

\author{\speaker{Emanuele R. Nocera}\\
        Rudolf Peierls Centre for Theoretical Physics, 
        University of Oxford,\\ 
        1 Keble Road, Oxford OX1 3NP, United Kingdom\\
        E-mail: \email{emanuele.nocera@physics.ox.ac.uk}}
\author{Huey-Wen Lin\\
        Department of Computational Mathematics, Science and Engineering, and\\
        Department of Physics and Astronomy,
        Michigan State University, East Lansing, MI 48824, USA}
\author{Fred Olness\\
        Department of Physics, Southern Methodist University, 
        Box 0175 Dallas, TX 75275, USA}
\author{Kostas Orginos\\
        Department of Physics, College of William and Mary, 
        Williamsburg, VA 23187-8795, and\\
        Jefferson Laboratory, 12000 Jefferson Avenue, Newport News, 
        VA 23606, USA}
\author{Juan Rojo\\
        Department of Physics and Astronomy, VU University, 
        NL-1081 HV Amsterdam, and\\
        Nikhef Theory Group, Science Park 105, 
        1098 XG Amsterdam, The Netherlands}

\abstract{The workshop on Parton Distributions and Lattice Calculations in the 
        LHC era (PDFLattice2017) was hosted at Balliol College, 
        Oxford (UK), from 22$^{\rm nd}$ to 24$^{\rm th}$ March 2017. 
        The workshop brought together the lattice-QCD and 
        the global-fit physicists who devote their efforts to 
        determine the parton distribution functions (PDFs) of the proton. 
        The goals were to make the two communities more familiar between each 
        other, review developments from both sides, and set precision 
        targets for lattice calculations so that they can contribute, 
        together with the forthcoming experimental input, to the next 
        generation of PDF determinations. This contribution summarises the 
        relevant outcome of the workshop, in anticipation of a thorough 
        white paper.}

\FullConference{XXV International Workshop on Deep-Inelastic Scattering 
                and Related Subjects\\
		3-7 April 2017\\
		University of Birmingham, UK}

\begin{document}

In Quantum Chromodynamics (QCD), Parton Distribution Functions (PDFs) 
are universal objects that encode the long-distance
dynamics of quarks and gluons interacting in a hard-scattering process.
Following factorisation~\cite{Collins:1989gx}, they 
are convoluted with partonic cross sections, that encode instead the 
short-distance dynamics of the interaction, in order to obtain predictions
for the experimental observables.
While partonic cross sections can be computed in QCD as a perturbative 
expansion in the strong coupling $\alpha_s$, PDFs cannot, although their
dependence on the factorisation scale $\mu$ results in the perturbatively
computable DGLAP evolution equations~\cite{Gribov:1972ri,
Altarelli:1977zs,Dokshitzer:1977sg}. 

The accurate knowledge of PDFs is at the base of the 
understanding of the nucleon structure, including how its momentum and spin 
are carried by quarks and gluons. 
Parton distributions are also fundamental tools in high-energy, nuclear and 
astroparticle phenomenology.
Their determination has thus received considerable theoretical and experimental 
attention over the years~\cite{Forte:2013wc,Jimenez-Delgado:2013sma}.

There are currently two main methods to determine PDFs.
The first method consists in a global QCD analysis of measurements of 
various hard-scattering observables.
Parton distributions are parametrised at an initial scale, evolved up to the
scale of the data, and used to build up the theoretical predictions for the 
relevant observables. 
In the corresponding factorisation formul\ae, the factorisation scale, $\mu$, 
is usually set equal to the characteristic scale of the process, $Q$.
The best-fit parameters are then determined by minimising a suitable figure of 
merit, such as the $\chi^2$.

Several theoretical and methodological details must be handled in a 
global QCD analysis.
On the theoretical side, general physical constraints must be enforced, 
specifically PDFs must lead to positive cross sections, and fulfill sum rules;
heavy quarks must be treated through an appropriate flavour scheme,
possibly extended to allow for intrinsic
components~\cite{Ball:2015dpa,Ball:2016neh}; and 
the highest-order QCD corrections should be included in the
evolution and in the computation of the partonic cross sections.
These are currently available at next-to-next-to-leading order (NNLO) 
in $\alpha_s$ for an increasing number of processes where the polarisation of 
the initial state is not measured, otherwise they are usually known only up to 
next-to-leading order (NLO).
As the precision of the measurements will increase, PDFs including Quantum 
Electrodynamics (QED) corrections~\cite{Ball:2013hta}, resummation 
effects~\cite{Bonvini:2015ira}, and higher-twist 
contributions~\cite{Accardi:2016qay,Sato:2016tuz} might be considered.

On the methodological side, particular attention is devoted to the 
determination of the PDF uncertainty, which is usually quantified 
with either the Hessian~\cite{Pumplin:2001ct} 
or the Monte Carlo~\cite{DelDebbio:2004xtd} method.
Both these methods allow one to account for various contributions to the 
PDF uncertainty: the measurement uncertainty propagated from the data, 
uncertainties associated with incompatibility of the fitted experiments, 
and procedural uncertainties such as those related to the choice of the PDF 
parametrisation.
Theoretical uncertainties, such as the parametric uncertainty due 
to the uncertainties on the values of the physical parameters used   
in the fit ({\it e.g.} the reference value of $\alpha_s$), and the missing 
higher order uncertainty (given that fits are performed with fixed-order 
perturbation theory), are instead more elusive.
While the size of the former can be estimated by varying the input parameters, 
the size of the latter is currently unknown, although it is supposed to be 
subdominant.

Several collaborations provide regular updates of PDF determinations 
from a global QCD analysis, both in the helicity-averaged (unpolarised, 
henceforth)~\cite{Accardi:2016qay,Harland-Lang:2014zoa,Dulat:2015mca,
Ball:2017nwa,Alekhin:2017kpj} and helicity-dependent (polarised, 
henceforth)~\cite{Sato:2016tuz,deFlorian:2014yva,Nocera:2014gqa} cases.
Differences in both the PDF central value and uncertainty from 
different sets are observed, mostly depending on the data set, 
and on the theoretical and methodological details of each QCD analysis.
In the unpolarised case, most of these differences~\cite{Alekhin:2011sk} are 
now understood, up to a point that some 
sets~\cite{Harland-Lang:2014zoa,Dulat:2015mca,Ball:2014uwa} were statistically 
combined into a single PDF set~\cite{Butterworth:2015oua}, 
with PDFs accurate to a few percent.
In the polarised case, instead, a more limited and less precise data set,
along with a lower degree in the theoretical and methodological sophistication
of the available QCD analyses, has led to less accurate PDFs.
This has also prevented from benchmarking various polarised PDF sets 
quantitatively so far.

The second method is provided by lattice QCD, {\it i.e.}, QCD formulated on a
finite-volume Euclidean spacetime discretised by means of the introduction of 
an ultraviolet cutoff.
Lattice QCD is generally studied by numerical computations of QCD correlation 
functions in the path-integral formalism, using methods adapted from 
statistical mechanics.
In oder to make contact with the data, numerical results have to be 
extrapolated to the continuum and infinite-volume limits.

Lattice-QCD calculations primarily determine the matrix elements of local 
twist-two operators that can be related to the Mellin moments of PDFs.
In principle, given a sufficient number of Mellin moments, PDFs can be 
reconstructed from the inverse Mellin transform.
In practice, calculations are limited to the lowest three 
moments~\cite{Alexandrou:2017oeh,Green:2012ud,Engelhardt:2012gd,
Gong:2015iir,Bratt:2010jn,Abdel-Rehim:2015owa,Bhattacharya:2016zcn,
Capitani:2017qpc}, because
power-divergent mixing occurs between twist-two operators.
Three moments are not enough to reconstruct the momentum fraction dependence 
of the PDFs without a significant model bias~\cite{Detmold:2003rq}. 
Novel strategies have been developed to compute higher 
moments~\cite{Davoudi:2012ya,Monahan:2015lha}, although they are still in 
their infancy.

Alternative methods have been proposed to determine the 
PDF momentum fraction dependence directly from lattice QCD,
among which the inversion method, the path-integral formulation
of the Deep-Inelastic Scattering (DIS) hadronic tensor, and quasi-PDFs.

The inversion method allows one to relate the unpolarised and polarised 
structure function $F_1$ and $g_1$ to the appropriate Compton amplitude through 
an integral equation, which can be solved numerically.
The Compton amplitude can be obtained by a simple 
extension~\cite{Chambers:2017dov} of existing implementations of the 
Feynamn-Hellman technique to lattice 
QCD~\cite{Horsley:2012pz,Chambers:2014qaa,Chambers:2015bka}.
Contributions from up, down and strange quarks, connected and 
disconnected, can be distinguished by appropriate insertions of 
the electromagnetic current.
The same method can be extended to PDFs, provided that $Q$ is sufficiently
large that power corrections in the Compton amplitude can be neglected.

The path-integral formulation of the DIS hadronic tensor~\cite{Liu:1993cv} 
provides an alternative formalism to carry out the operator product expansion.
There are three gauge-invariant and topologically distinct contributions
to the hadronic tensor - respectively from the valence and the connected 
and disconnected sea quarks - which can be computed on the Euclidean lattice 
by evaluation of four-point functions. 
The hadronic tensor should be converted from the Euclidean to the 
Minkowski space (for details on numerical approaches, 
see~\cite{Liu:2016djw}).
Once it is extended to the continuum and large volume limits and at the 
physical pion mass, one can apply QCD factorisation to fit the PDFs.

Quasi-PDFs~\cite{Ji:2013dva} are defined as appropriate 
momentum-dependent nonlocal static matrix elements for nucleon states at
finite momentum, with an ultraviolet cut-off scale, {\it e.g.} the inverse of 
the lattice spacing.
Quasi-PDFs must be related to the corresponding light-front PDFs, for which
the nucleon momentum is taken to infinity.
This is usually achieved by means of a matching kernel in the Large-Momentum 
Effective field Theory (LaMET)~\cite{Ji:2013dva}.
Approaches alternative to LaMET~\cite{Ma:2014jla} view quasi-PDFs as a 
{\it lattice cross section} from which the light-front PDF can be factorised. 
Related constructions were proposed in~\cite{Radyushkin:2016hsy} and explored 
in~\cite{Orginos:2017kos}.
A procedure combining information from quasi-PDFs and moments of PDFs
was tested in~\cite{Bacchetta:2016zjm}.
Lattice calculations of quasi-PDFs have been ecouraging so far, 
although they are still rather qualitative~\cite{Lin:2014zya,
Alexandrou:2015rja,Chen:2016utp}.

In order to make meaningful contact with the data, lattice QCD calculations 
must demonstrate control over a wide range of systematic uncertainties 
introduced by the discretisation of QCD on the lattice.
These include discretisation effects that vanish in the continuum limit,
extrapolation from unphysically pion masses, finite volume effects,
excited state contamination, and renormalisation of composite operators.
The continuum limit also requires an accurate determination of the 
lattice spacing, which however introduces a negligible uncertainty.
All these sources of systematics are critically reviewed in~\cite{Aoki:2016frl}.
In addition, quasi-PDFs are subject to uncertainties associated with the
finite nucleon momentum of the lattice calculation, and with their specific
renormalisation.

The workshop on Parton Distributions and Lattice Calculations in the LHC era
(PDFLattice2017)~\cite{workshop} was organised to bring together 
physicists who actively work to determine PDFs either 
from global fits or from lattice QCD.
It was hosted at Balliol College, Oxford (UK), from 22$^{\rm nd}$ to
24$^{\rm th}$ March 2017.
The goals were to make the global-fit and lattice-QCD communities more familiar
between each other, review recent developments from both sides, and
discuss how lattice-QCD calculations can be used to 
improve global fits, and, conversely, how global fits can be used to benchmark
lattice-QCD calculations. 
The workshop was specifically focused on precision physics, and
included aspects of both the high-energy physics program at the Large Hadron 
Collider (LHC), and the hadron physics program at the Relativistic Heavy Ion Collider (RHIC), at Jefferson Lab (JLab), and at other facilities.
Therefore, the discussion was consciously limited to collinear unpolarised and 
polarised PDFs.
Future editions of the workshop could be extended to transversity,
Transverse-Momentum-Dependent PDFs (TMDs), and Generalised PDFs (GPDs).

\begin{figure}[!t]
\centering
\begin{tikzpicture}[node distance = 1.4cm, auto]
 \footnotesize
 \node [block, left=1cm] (pQCD) {Global QCD analyses\\ PDF fits};
 \node [block, right=1cm] (lQCD) {Lattice QCD\\ moments/quasi-PDFs};
 \path [line, thick] (pQCD) |- (0,+1) -| (lQCD);
 \path [line, dashed, thick] (+3,0.5) |-++ (0,+1) --++ (-6,0) -|++ (0,-1);
 \node at (-1,+1) [above=2mm, right=0mm] {\textsc{\footnotesize{benchmark}}}; 
 \path [line, thick] (lQCD) |- (0,-1) -| (pQCD);
 \path [line, dashed, thick] (-3,-0.5) |-++ (0,-1) --++ (6,0) -| (3,-0.5);
 \node at (-0.6,-1) [below=2mm, right=0mm] {\textsc{\footnotesize{input}}}; 
 \node [below of=pQCD, node distance=2cm]  (text11) {\underline{LHC (precision physics)}};
 \node [below of=text11, node distance=0.5cm] (text21) {Higgs boson characterisation};
 \node [below of=text21, node distance=0.4cm] (text31) {Precision SM measurements ({\it e.g.} $M_W$)};
 \node [below of=text31, node distance=0.4cm] (text41) {BSM searches, SUSY};   
 \node [below of=lQCD, node distance=2cm]  (text12) {\underline{RHIC, JLab, \dots (hadron physics)}};
 \node [below of=text12, node distance=0.5cm] (text22) {Spin physics, nucleon structure};
 \node [below of=text22, node distance=0.4cm] (text32) {Large-$x$ behaviour};
 \node [below of=text32, node distance=0.4cm] (text42) {Nuclear modifications};
 \draw [thick,dashed](-7.36,-4.3) rectangle (+7.36,1.8);
 \node at (-7.36,-4.3) [above=2mm, right=0mm] {\textsc{Global QCD fit and Lattice QCD interplay in PDF determinations}};   
 \node at(0,-5.0) {Define a mutually agreed conventional notation for relevant PDF-related quantities, such as PDF moments.};
 \node at(0,-5.5) {Assess the sources of systematic uncertainties in lattice-QCD calculations.};
 \node at(0,-6.0) {Identify a best-set of quantities to benchmark lattice-QCD calculations against global-fit determinations.};
 \node at(0,-6.5) {Set precision targets for lattice-QCD calculations with respect to global-fit determinations.};
 \node at(0,-7.0) {Assess the impact of lattice-QCD calculations on global-fit determinations within their current/projected precision.};
 \draw [thick,dashed](-7.36,-7.7) rectangle (+7.36,-4.5);
 \node at (-7.36,-7.7) [above=2mm,right=0mm] {\textsc{Desiderata}}; 
 \draw [thick](-7.48,-7.82) rectangle (+7.48,2.5);
 \node at (-7.36,2.5) [below=4mm,right=0mm] {\textsc{\large The PDFLattice2017 workshop}}; 
\end{tikzpicture}
\caption{A graphical summary of the PDFLattice2017 workshop outcome.}
\label{fig:outcome}
\end{figure}
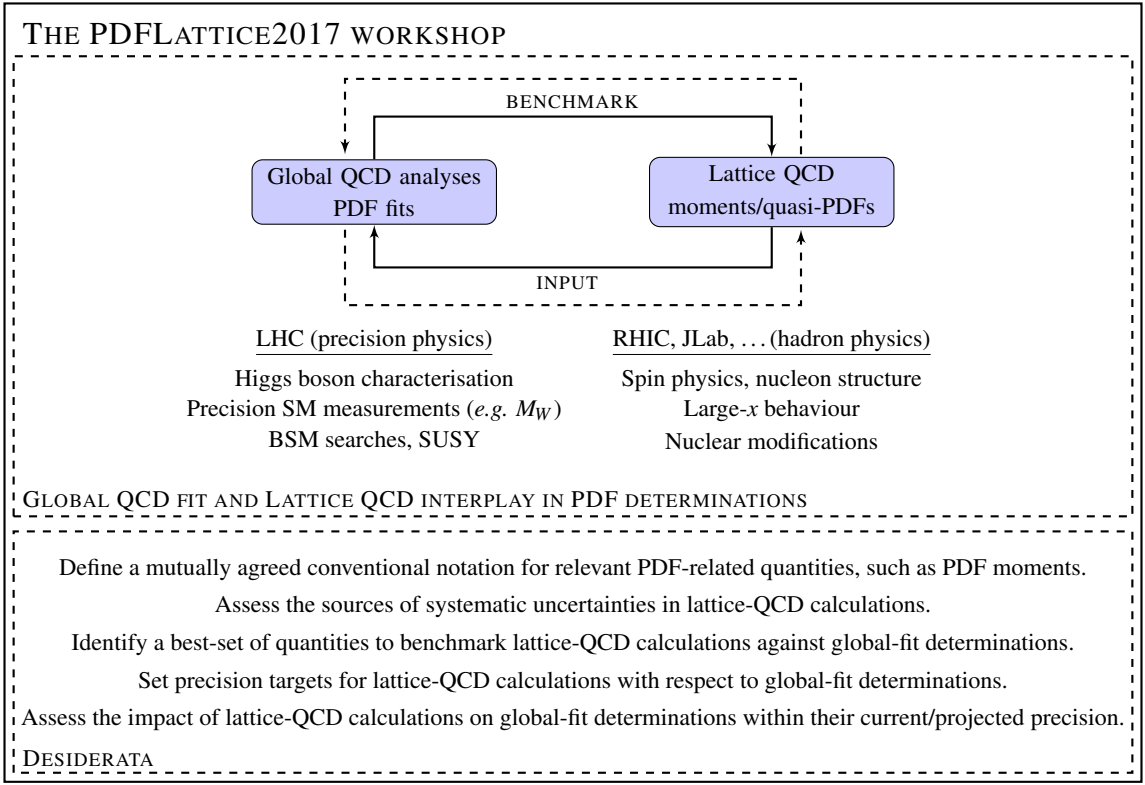

A graphical summary of the workshop outcome is displayed in 
Fig.~\ref{fig:outcome}.
It includes the following list of {\it desiderata}, which was compiled in order 
to strengthen the cross-talk between the two PDF communities in the future.
First, define a common language, including a mutually agreed conventional 
notation for relevant PDF-related quantities, such as PDF moments.
Second, assess the sources of systematic uncertainties in lattice-QCD
calculations, {\it e.g.} along the lines of~\cite{Aoki:2016frl}.
Third, indentify a set of quantities to benchmark lattice-QCD results against
global-fit determinations, and provide a prescription to select and/or 
combine the most reliable and up-to-date results from both sides.
Fourth, set precision targets for lattice-QCD calculations so that they
can contribute, together with the forthcoming experimental input, to the
next generation of PDF determinations.
Fifth and last, assess the impact of lattice-QCD calculations on global-fit 
determinations assuming their current and/or projected precision.

All these points will be addressed thoroughly in a forthcoming white 
paper~\cite{whitepaper}.
In particular, this will include a full review of recent developments in 
lattice-QCD and global-fit PDF determinations, a 
complete set of up-to-date benchmark numbers for the 
relevant moments from both sides, and an assessment of the impact of 
lattice-QCD calculations on global-fit determinations assuming their 
current and/or projected precision.
The white paper will represent the joint effort of the global-fit and
lattice-QCD PDF communities to make the most of the lively activities
spurred by the workshop.
It will hopefully become a reference and a further motivation to encourage 
fruitful interactions between the two communities.


I would like to thank all the participants to the PDFLattice2017 workshop
for making it very fruitful.
I am also grateful to Jacqueline Gills and Michelle Bosher for their help
in the organisation of the workshop.
The workshop was partly supported by the European Research Council via the 
Starting Grant {\it PDF4BSM - Parton Distributions in the Higgs Boson Era}.
I acknowledge financial support from the UK STFC Rutherford Grant ST/M003787/1.


\begin{thebibliography}{99}

\bibitem{Collins:1989gx}
  J.~C.~Collins, D.~E.~Soper and G.~F.~Sterman,
  Adv.\ Ser.\ Direct.\ High Energy Phys.\  {\bf 5} (1989) 1.

\bibitem{Gribov:1972ri}
  V.~N.~Gribov and L.~N.~Lipatov,
  Sov.\ J.\ Nucl.\ Phys.\  {\bf 15} (1972) 438
  [Yad.\ Fiz.\  {\bf 15} (1972) 781].

\bibitem{Altarelli:1977zs}
  G.~Altarelli and G.~Parisi,
  Nucl.\ Phys.\ B {\bf 126} (1977) 298.

\bibitem{Dokshitzer:1977sg}
  Y.~L.~Dokshitzer,
  Sov.\ Phys.\ JETP {\bf 46} (1977) 641
  [Zh.\ Eksp.\ Teor.\ Fiz.\  {\bf 73} (1977) 1216].

\bibitem{Forte:2013wc}
  S.~Forte and G.~Watt,
  Ann.\ Rev.\ Nucl.\ Part.\ Sci.\  {\bf 63} (2013) 291.

\bibitem{Jimenez-Delgado:2013sma}
  P.~Jimenez-Delgado, W.~Melnitchouk and J.~F.~Owens,
  J.\ Phys.\ G {\bf 40} (2013) 093102.

\bibitem{Ball:2015dpa}
  R.~D.~Ball {\it et al.},
  JHEP {\bf 1511} (2015) 122;
  Phys.\ Lett.\ B {\bf 754} (2016) 49.

\bibitem{Ball:2016neh}
  R.~D.~Ball {\it et al.} [NNPDF Collaboration],
  Eur.\ Phys.\ J.\ C {\bf 76} (2016) 647.

\bibitem{Ball:2013hta}
  R.~D.~Ball {\it et al.} [NNPDF Collaboration],
  Nucl.\ Phys.\ B {\bf 877} (2013) 290.

\bibitem{Bonvini:2015ira}
  M.~Bonvini {\it et al.},
  JHEP {\bf 1509} (2015) 191;
  L.~Rottoli and M.~Bonvini,
  arXiv:1707.01535 [hep-ph].

\bibitem{Accardi:2016qay}
  A.~Accardi {\it et al.}
  Phys.\ Rev.\ D {\bf 93} (2016) 114017.

\bibitem{Sato:2016tuz}
  N.~Sato {\it et al.} [JAM Collaboration],
  Phys.\ Rev.\ D {\bf 93} (2016) 074005.

\bibitem{Pumplin:2001ct}
  J.~Pumplin {\it et al.}
  Phys.\ Rev.\ D {\bf 65} (2001) 014013.

\bibitem{DelDebbio:2004xtd}
  L.~Del Debbio {\it et al.} [NNPDF Collaboration],
  JHEP {\bf 0503} (2005) 080.

\bibitem{Harland-Lang:2014zoa}
  L.~A.~Harland-Lang {\it et al.},
  Eur.\ Phys.\ J.\ C {\bf 75} (2015) 204.

\bibitem{Dulat:2015mca}
  S.~Dulat {\it et al.},
  Phys.\ Rev.\ D {\bf 93} (2016) 033006.

\bibitem{Ball:2017nwa}
  R.~D.~Ball {\it et al.} [NNPDF Collaboration],
  arXiv:1706.00428 [hep-ph].

\bibitem{Alekhin:2017kpj}
  S.~Alekhin, J.~Blümlein, S.~Moch and R.~Placakyte,
  Phys.\ Rev.\ D {\bf 96} (2017) 014011.

\bibitem{Ball:2014uwa}
  R.~D.~Ball {\it et al.} [NNPDF Collaboration],
  JHEP {\bf 1504} (2015) 040.

\bibitem{deFlorian:2014yva}
  D.~de Florian, R.~Sassot, M.~Stratmann and W.~Vogelsang,
  Phys.\ Rev.\ Lett.\  {\bf 113} (2014) 012001.

\bibitem{Nocera:2014gqa}
  E.~R.~Nocera {\it et al.} [NNPDF Collaboration],
  Nucl.\ Phys.\ B {\bf 887} (2014) 276.

\bibitem{Alekhin:2011sk}
  S.~Alekhin {\it et al.},
  arXiv:1101.0536 [hep-ph];
  R.~D.~Ball {\it et al.},
  JHEP {\bf 1304} (2013) 125.

\bibitem{Butterworth:2015oua}
  J.~Butterworth {\it et al.},
  J.\ Phys.\ G {\bf 43} (2016) 023001.

\bibitem{Alexandrou:2017oeh}
  C.~Alexandrou {\it et al.},
  arXiv:1706.02973 [hep-lat].

\bibitem{Green:2012ud}
  J.~R.~Green {\it et al.},
  Phys.\ Lett.\ B {\bf 734} (2014) 290;
  G.~S.~Bali {\it et al.},
  Phys.\ Rev.\ D {\bf 90} (2014) 074510.

\bibitem{Engelhardt:2012gd}
  M.~Engelhardt,
  Phys.\ Rev.\ D {\bf 86} (2012) 114510;
  Y.~Aoki {\it et. al.},
  Phys.\ Rev.\ D {\bf 82} (2010) 014501.

\bibitem{Gong:2015iir}
  M.~Gong {\it et al.} [$\chi$QCD Collaboration],
  Phys.\ Rev.\ D {\bf 95} (2017) 114509.

\bibitem{Bratt:2010jn}
  J.~D.~Bratt {\it et al.} [LHPC Collaboration],
  Phys.\ Rev.\ D {\bf 82} (2010) 094502.

\bibitem{Abdel-Rehim:2015owa}
  A.~Abdel-Rehim {\it et al.},
  Phys.\ Rev.\ D {\bf 92} (2015) 114513;
  Erratum: [Phys.\ Rev.\ D {\bf 93} (2016) 039904].

\bibitem{Bhattacharya:2016zcn}
  T.~Bhattacharya {\it et al.},
  Phys.\ Rev.\ D {\bf 94} (2016) 054508.

\bibitem{Capitani:2017qpc}
  S.~Capitani {\it et al.},
  arXiv:1705.06186 [hep-lat];
  G.~S.~Bali {\it et al.},
  Phys.\ Rev.\ D {\bf 91} (2015) 054501.

\bibitem{Detmold:2003rq}
  W.~Detmold, W.~Melnitchouk and A.~W.~Thomas,
  Mod.\ Phys.\ Lett.\ A {\bf 18} (2003) 2681.

\bibitem{Davoudi:2012ya}
  Z.~Davoudi and M.~J.~Savage,
  Phys.\ Rev.\ D {\bf 86} (2012) 054505.

\bibitem{Monahan:2015lha}
  C.~Monahan and K.~Orginos,
  Phys.\ Rev.\ D {\bf 91} (2015) 074513.

\bibitem{Chambers:2017dov}
  A.~J.~Chambers {\it et al.},
  Phys.\ Rev.\ Lett.\  {\bf 118} (2017) 242001.

\bibitem{Liu:1993cv} 
  K.~F.~Liu and S.~J.~Dong,
  Phys.\ Rev.\ Lett.\  {\bf 72}, 1790 (1994);
  K.~F.~Liu,
  Phys.\ Rev.\ D {\bf 62}, 074501 (2000).

\bibitem{Liu:2016djw} 
  K.~F.~Liu,
  PoS LATTICE {\bf 2015}, 115 (2016).

\bibitem{Horsley:2012pz}
  R.~Horsley {\it et al.} [QCDSF and UKQCD Collaborations],
  Phys.\ Lett.\ B {\bf 714} (2012) 312.

\bibitem{Chambers:2014qaa}
  A.~J.~Chambers {\it et al.} [CSSM and QCDSF/UKQCD Collaborations],
  Phys.\ Rev.\ D {\bf 90} (2014) 014510.

\bibitem{Chambers:2015bka}
  A.~J.~Chambers {\it et al.},
  Phys.\ Rev.\ D {\bf 92} (2015) 114517.

\bibitem{Ji:2013dva}
  X.~Ji,
  Phys.\ Rev.\ Lett.\  {\bf 110} (2013) 262002;
  Sci.\ China Phys.\ Mech.\ Astron.\  {\bf 57} (2014) 1407.

\bibitem{Ma:2014jla}
  Y.~Q.~Ma and J.~W.~Qiu,
  arXiv:1404.6860 [hep-ph];
  Int.\ J.\ Mod.\ Phys.\ Conf.\ Ser.\  {\bf 37} (2015) 1560041.

\bibitem{Radyushkin:2016hsy}
  A.~Radyushkin,
  Phys.\ Lett.\ B {\bf 767} (2017) 314;
  arXiv:1705.01488 [hep-ph].

\bibitem{Orginos:2017kos}
  K.~Orginos, A.~Radyushkin, J.~Karpie and S.~Zafeiropoulos,
  arXiv:1706.05373 [hep-ph].

\bibitem{Bacchetta:2016zjm}
  A.~Bacchetta, M.~Radici, B.~Pasquini and X.~Xiong,
  Phys.\ Rev.\ D {\bf 95} (2017) 014036.

\bibitem{Lin:2014zya}
  H.~W.~Lin {\it et al.},
  Phys.\ Rev.\ D {\bf 91} (2015) 054510;
  H.~W.~Lin {\it et al.},
  arXiv:1708.05301 [hep-lat].

\bibitem{Alexandrou:2015rja}
  C.~Alexandrou {\it et al.},
  Phys.\ Rev.\ D {\bf 92} (2015) 014502;
  Phys.\ Rev.\ D {\bf 96} (2017) 014513.

\bibitem{Chen:2016utp}
  J.~W.~Chen {\it et al.},
  Nucl.\ Phys.\ B {\bf 911} (2016) 246;
  arXiv:1706.01295 [hep-lat].

\bibitem{Aoki:2016frl}
  S.~Aoki {\it et al.},
  Eur.\ Phys.\ J.\ C {\bf 77} (2017) 112.

\bibitem{workshop}
  See the workshop web page~\href{http://www.physics.ox.ac.uk/confs/PDFlattice2017/index.asp}{http://www.physics.ox.ac.uk/confs/PDFlattice2017/index.asp}.

\bibitem{whitepaper}
  H.-W. Lin, E.~R. Nocera, F.~Olness, K.~Orginos, J.~Rojo (editors)
  {\it et al.}, in preparation.

\end{thebibliography}
\end{document}